# Shifting Policy Strategy in Keynesianism[1]

## Asahi Noguchi[2]

## Abstract

*This paper analyzes the evolution of Keynesianism making use of concepts offered by Imre Lakatos. The Keynesian "hard core" lies in its views regarding the instability of the market economy, its "protective belt" in the policy strategy for macroeconomic stabilization using fiscal policy and monetary policy. Keynesianism developed as a policy program to counter classical liberalism, which attributes priority to the autonomy of the market economy and tries to limit the role of government. In general, the core of every policy program consists in an unfalsifiable worldview and a value judgment that remain unchanged. On the other hand, a policy strategy with a protective belt inevitably evolves owing to changes in reality and advances in scientific knowledge. This is why the Keynesian policy strategy has shifted from being fiscal-led to one that is monetary-led because of the influence of monetarism; further, the Great Recession has even led to their integration.*

# I. Introduction

Keynesianism is a policy ideology that prescribes an active anti-cyclical government policy to stabilize an inherently unstable market economy, thereby achieving the desired levels of employment and income. Keynesianism was proposed as an antithesis to the classical laissez-faire doctrine, under which a government does nothing to impede the autonomous adjustment mechanism of a market. This conceptual core of Keynesianism has not changed much since it was originally established. However, the policy strategy for attaining the desired employment and income under Keynesianism has changed over time.

---

[1] This work was supported by JSPS KAKENHI Grant Number JP16K03578 "Theory and Thought of Macroeconomic Policy in Economic Crisis".
[2] Senshu University, School of Economics, Tokyo, Japan

Most early Keynesians considered fiscal policy to be the main axis of macroeconomic policy, with monetary policy seen as playing a supplementary role. This view was based on skepticism about monetary policy and the recognition that the interest rate is money inelastic, while investment is interest-rate inelastic. By contrast, the early Keynesians had strong confidence in fiscal policy; this arose partly from their belief in the fiscal multiplier theory.

This fiscal-led Keynesianism had largely lost force owing to the criticism of Keynesian economics by Milton Friedman and the subsequent "monetarist counter-revolution." The emergence of high rates of inflation, beginning in the late 1960s, followed by stagflation in the 1970s, contributed to this shift from Keynesianism to monetarism. Although there are exceptions, fiscal policy had rarely been used as a stimulus measure since the 1970s. Under the influence of monetarism, the policy strategy in Keynesianism also shifted from fiscal to monetary policy. The result was that macroeconomic stabilization had been left almost exclusively to monetary policy. The era of the Great Moderation, from the latter half of the 1980s to the first half of 2000s, represented the acme of this monetary-led Keynesianism.

Owing to the global financial crisis in 2008 and the subsequent Great Recession, policy strategy along the lines of Keynesianism underwent further transformation. It first changed from traditional monetary policy to the full utilization of non-traditional monetary policy. Further, it was accompanied by the integration of monetary and fiscal policies. The latter is also seen as a strategy against the global "austerity" imposed in response to the European sovereign debt crisis, which began in the spring of 2010. This flexibility of Keynesian policy strategy is the main reason for its amazing vitality as a policy program.

## II. Why Didn't Keynesianism Die?

It was at the end of the 1970s that Robert Lucas declared "the death of Keynesian economics" in his essay for graduate students at the University of Chicago; it began:

> The main development I want to discuss has already occurred: Keynesian economics is dead [maybe 'disappeared' is a better term]. I do not exactly know when this happened but it is true today and it was not true two years ago. This is a sociological not an economic observation, so evidence for it is sociological. For example, one cannot find a good, under 40 economist who identifies himself, works as 'Keynesian'. Indeed, people even take offence if referred to in this way. At research seminars, people do not take Keynesian theorizing seriously any more — audience starts to whisper and giggle to one another. Leading journals are not getting Keynesian papers submitted any more.

I suppose I, and with many others, were in on the kill in an intellectual sense, but I do not say this as any kind of boast, or even with much pleasure. Just a fact. (Lucas 2013, 500-501)

It is apparent, at least in the academic world of economics, that this "sociological observation" of Lucas was neither a boast nor self-promotion. If a graduate student wanted to conduct research in the field of macroeconomics, he or she would usually be required to ignore Keynesian economics and follow, instead, the "new classical" macroeconomics developed by Lucas and other anti-Keynesian economists.

However, Lucas' prediction did not come true. Following the above victory declaration, Lucas stated:

True, there are still leading Keynesians — in academics and government circles — so Keynesian economics is alive in this sense — but this is transient, because there is no fresh source of supply. The only way to produce a 60 year old Keynesian is to produce a 30 year old Keynesian, and wait 30 years. So the implications for policy will take a while to be evident — but they can be very accurately predicted. (Lucas 2013, 501)

The subsequent turn of events belied this prediction in two ways. First, the idea of Lucas and the others certainly revolutionized macroeconomics, but did not exterminate the "30 year old Keynesian." In fact, a new generation of Keynesians, who came to be called the new Keynesians, appeared. Rather than abandoning Keynesian ideas and converting to the new classical school, they began to build the new Keynesian economics that incorporated ideas and analytical tools developed by the new classical economists. Second, even 30 or 40 years after Lucas' prediction, the new classical economists were never able to rule the policy world by displacing Keynesian economists.

In his 2006 article titled "The Macroeconomist as Scientist and Engineer," Gregory Mankiw, who is one of the representatives of new Keynesian economics, stated that:

Among the leaders of the new classical school, none (as far as I know) has ever left academia to take a significant job in public policy. By contrast, the new Keynesian movement, like the earlier generation of Keynesians, was filled with people who traded a few years in the ivory tower for a stay in the nation's capital. Examples include Stanley Fischer, Larry Summers, Joseph Stiglitz, Janet Yellen, John Taylor, Richard Clarida, Ben Bernanke, and myself. The first four of these economists came to Washington during the Clinton years; the last four during the

> Bush years. The division of economists between new classicals and new Keynesians is not, fundamentally, between the political right and the political left. To a greater extent, it is a split between pure scientists and economic engineers. (Mankiw 2006, 37)

On this evidence, it does seem that the policy world is, after all, occupied almost exclusively by Keynesians, as was the case earlier.

As the title of Mankiw's paper suggests, he is likening the new classicals and new Keynesians to pure scientists and engineers, respectively. However, such a characterization is not necessarily appropriate. As is evident in the case of monetarism, there clearly existed a policy program and strategy based on classical thinking. It was the application to society of a doctrine that was derived from a certain value judgment, thereby taking on an obvious engineering aspect. The reality is that the social engineering of the classicals has disappeared, while that of the Keynesians has survived in the area of macroeconomic policy. The essential question to ask is why this happened.

Keynesianism has shown an astonishing vitality in the policy world. This is because Keynesianism, at the level of policy program, is "progressive" in the sense that Imre Lakatos meant in the context of a scientific research program (Lakatos 1970). The progressiveness of a policy program means that it can continue to produce a new policy strategy that is applicable to whatever changes that occur in the real world, while maintaining its intrinsic worldview and value judgment. In a policy program, the worldview and value judgments are always at its hard core, and a policy strategy that can evolve with changing reality and newer knowledge is at its protective belt. The Keynesianism policy program is apparently progressive in this sense, for it continues to update its policy strategy in line with changing reality and newer knowledge. Such practicality and flexibility in Keynesianism is exactly what classical anti-Keynesian policy programs sorely lacked.

In fact, Keynesian policy strategies have continued to change dramatically over time. The new classical macroeconomics emerged from the criticism of Keynesian economics. Naturally, Keynesians were initially antagonistic toward it. Before long, however, some of them began to accept the criticism of its weakness. They then began the endeavor to overcome its weakness by incorporating these new classical ideas and theories into the Keynesian policy program. As a result, the protective belt in Keynesian policy program has been constantly strengthened. As is shown in the following sections, Keynesian policy strategy evolved from an early fiscal-led Keynesianism to a monetary-led one, that is, from Keynesianism I to Keynesianism II. In recent years, Keynesianism II has undergone further transformation to achieve closer integration between deficit fiscal policy and monetary policy in its ideological battle against the austerity doctrine.

# III. Historical Roles and Problems in Fiscal-Led Keynesianism

1. Why Did the Early Keynesians Emphasize Fiscal Policy and not Monetary Policy?

Keynesianism, when it initially found its way into policymaking, was dominated by fiscal policy. The impression is so strong that the notion prevails even today that a Keynesian policy is no more or less than an economic measure involving extended fiscal spending. However, Keynes' own grasp of policy went much beyond fiscal spending. Had it not been the case, Keynes would not have gone into detailed discussion of "money" and "interest" in his *General Theory* (1936). Further, he would have been satisfied with simply presenting an income determination theory based on aggregate demand analysis centered on the fiscal multiplier theory, which we can see in Chapter 10 of his *General Theory*.

In contrast, it is clear that many of the early Keynesians, who inherited Keynes's thought, were skeptical about the effectiveness of monetary policy. This is because the Keynesians of that time thought that the effects of monetary policy were confined to a single channel, whereby a decline in interest rates would lead to an expansion of private investment. Contemporary economists know that monetary policy works on the real economy not only through the channel of interest rates, but also through other channels, such as exchange rates, asset prices (including stock prices), credit, expectations, and so on. However, the Keynesians of the time did not have this understanding.

It is clear that monetary policy works on aggregate demand in a more indirect way than does fiscal policy. Further, the effectiveness of the interest rate channel seems very limited because it depends on the reduction in interest rate caused by the monetary expansion, as well as the expansion of private investment resulting from it. The elasticity of interest rate to money supply and that of investment to interest rate may not be enough.

First, the interest rate is clearly not always elastic to the money supply. According to Keynes, interest is the price paid to relinquish liquidity. Therefore, when interest rates are low, people tend to prefer money – i.e. the most liquid asset. This means that the decline in interest rates becomes ever more moderate as the money supply increases. In other words, interest rates gradually become more inelastic to the money supply. When the interest rate is very low, it becomes completely inelastic. This is the "liquidity trap" that Keynes pointed out in Chapter 15 of *General Theory*. In a liquidity trap, the interest rate channel of monetary policy no longer works.

Further, even if there is room for a decrease in the interest rate that is sufficiently elastic to the money supply, it is not always clear whether private investment would expand sufficiently as interest rates are reduced. A pessimistic finding regarding this is the famous Oxford Survey, which was conducted by an economic research group at Oxford University in the 1930s. According to the survey, firms' investment decisions are hardly influenced by interest rates at all.[3] If this finding holds universally, it leads to the conclusion that the interest rate channel of monetary expansion cannot reliably lead to higher private investment, regardless of whether the economy is in a liquidity trap. The early Keynesians largely shared this understanding about the interest rate channel in monetary policy, which was often referred to as "elasticity pessimism".

On the other hand, the demand expansion effect of fiscal policy is obviously much more direct than monetary policy. In particular, government expenditure, such as public investment, leads to a direct increase in public demand. If fiscal policy is implemented through tax cuts or pecuniary benefits to households, it does not necessarily lead to a corresponding increase in private demand because there is a possibility that households will divert some of the increase in disposable income to savings, rather than consumption. Even in that case, if these public expenditures were financed by government deficit bonds instead of taxes, the increase in disposable income of households would lead to more consumption, which would in turn lead to an overall increase in demand in the economy.

According to Keynesian economics, fiscal policy extends well beyond such direct spending by government and households as are covered by government bonds. It also has an expansive ripple effect through the chain of income and demand, which amplifies the demand and income of the economy. This is the multiplier theory that first appeared in Chapter 10 of the *General Theory* titled "The Marginal Propensity to Consume and the Multiplier". Paul Samuelson reformulated the theory as the 45-degree line model, which became popular among the general public as a synonym for Keynesian economics. This 45-degree line model or the fiscal multiplier theory is the basic theory that supports Keynesianism I, or fiscal-led Keynesianism.

2. Significance of, and Problems in, Samuelson's 45-Degree Line Model

The 45-degree line model first appeared in Samuelson's *Economics*, the first edition of which was published in 1948 (Samuelson 1997). The textbook achieved great commercial success. As a result, the 45-degree line model, which is the simplest Keynesian model, and the concept of the fiscal multiplier based on it, have become common knowledge among the general public with some economic literacy. In that sense, Samuelson's *Economics* played an important role as a Keynesian evangelist.

Samuelson's 45-degree model is still frequently used in introductory classes in macroeconomics. It is certainly useful as an educational tool to show the interdependence of income and expenditure in the macroeconomy – that is, how someone's income is the result of someone else's spending. However, the model is rarely used as a tool by professional economists to conduct more formal research on policy analysis. This is because its primary conclusion about the government's fiscal spending bringing about a multifold increase in demand is hardly true in reality.

---

[3] On this point, see Meade and Andrews (1952, 28-30)

The biggest problem with the 45-degree line model is its basic premise that people always use a certain percentage of their income for consumption, which was later called the Keynesian consumption function. Apparently, people's actual consumption and savings behavior is not so simple; it depends on disposable income over longer time horizons. Instead of relying only on current income, it includes assets that have been saved so far and income that could be earned in the future.

With these problems in mind, after Keynes economists tried to replace his treatment of consumption with something more realistic. The results were the permanent income hypothesis by Milton Friedman and the life cycle hypothesis by Franco Modigliani.[4] The permanent income hypothesis proposes that our consumption depends not only on the current income, as Keynes assumes, but also on the permanent income, i.e. the expected long-term average income. The life cycle hypothesis proposes that our consumption depends on the lifetime income that each individual can earn in the course of his or her life. Both these hypotheses are applications of the fundamental economic principle that individuals maximize utility under the income constraints they face and choose to consume or save within their lifetime.

The primary policy implication derived from Keynesian macroeconomic theory — the multiplier effect of the fiscal stimulus — depends on this very special setting of the Keynesian consumption function. In the multiplier theory, the ripple effect of consumption occurs because people are always assumed to divert a part of their increase in income to consumption. If people divert all their income increases to savings, rather than consumption, there is no consumption ripple effect at all. Of course, there are always some households that immediately consume the increase in income. However, this would probably be limited to poor households with few assets. In a typical household, with a certain amount of assets, the magnitude of consumption at each point in time would not necessarily depend on the current income. Considering these problems, it was clear from the outset that the fiscal multiplier theory based on the 45-degree line model should have been treated exclusively as an educational model rather than the reference model for actual macroeconomic policy.

3. Demise of Fiscal-led Keynesianism

Keynesianism I or fiscal-led Keynesianism calls for fiscal policy as a primal means of macroeconomic stabilization; it reached the height of its influence in the policy world in the first half of the 1960s. In Democrat John F. Kennedy's administration, which took office in 1961, leading Keynesian economists, such as James Tobin and Robert Solow, acted as

---

[4] The permanent income hypothesis and the life cycle hypothesis first appeared in Friedman (1957) and

policy advisers. The policy program they conducted was genuinely in the spirit of Keynesianism, and its policy goal of achieving full employment was clear evidence of this. Naturally, fiscal policy was assigned as the main means for achieving that goal. The theory related to this economic policy was termed the "New Economics".

The fiscal policy actually applied by the Kennedy administration was income tax reduction, known as the Kennedy tax cut, which was proposed by the Kennedy administration and implemented in 1964, after Kennedy's assassination. In fact, many Keynesian economists in the administration advised Kennedy that fiscal expenditure was preferable to tax reduction for economic recovery. However, it was the tax reduction policy that Kennedy actually chose because he thought that the maximum US income tax rate of 91% at that time was too high.

The policy strategy characterized by Keynesianism I rapidly disappeared from the policy world at the end of the 1960s. This was mainly because monetarism, as an anti-Keynesian policy program, penetrated academia and the policy world around that time. The economic condition of the developed economies during this period, that is, high inflation as from the end of the 1960s and stagflation as from the 1970s, motivated the anti-Keynesian movement. Around 1980, the Thatcher administration in the United Kingdom and the Reagan administration in the US took office one after the other. This marked a major turning point in the history of economic policy, in that monetarism in the macroeconomic sphere and neoliberalism in the microeconomic sphere penetrated the real policy world.

What must not be overlooked is the fact that the demise of Keynesianism I was already in progress within academia before monetarism began to influence the policy world. Even Keynesian economists began to avoid using bare fiscal multiplier theory relying on the Keynesian-type consumption function that lacks a micro foundation. Further, the dogma of preferring fiscal policy over everything derived from the former approach had often come to be viewed as "crude Keynesianism". Robert Lucas' "sociological observation" that attendees at economics seminars started to ignore Keynesian types of studies had already been noticed long before he wrote about it.

In fact, the thinking behind Keynesianism I concepts came to be criticized in the late 1960s not only by the anti-Keynesian monetarists, but also by some Keynesians. Axel Leijonhufvud, who was apparently one of the most influential adherents to the Keynesian camp at that time, stated:

> The discussion of the interest-elasticity of investment has proceeded within the debate over the efficacy of monetary policies in combating business fluctuations. Keynes' pessimism on the latter issue and his propaganda for various fiscal policies were two of the most prominent features of the *General Theory*. In the hands of many early "Keynesians", these elements of his thought hardened into simplified

> dogmas — monetary policy came to be regarded as completely ineffective in recession while fiscal policies were propounded as the universal and only cure for macroeconomic problems. In the course of this evolution, there was an important change of emphasis in the explanation offered for the presumed inefficacy of monetary measures. It is this shift of emphasis which concerns us here. (Leijonhufvud 1968,158)

In short, Leijonhufvud pointed out that fiscal-led Keynesianism, which was characteristic of early Keynesians, should be abandoned if Keynes' original thinking was to be respected.

## IV. From Monetarist Counter-Revolution to Keynesianism II

### 1. Monetarism as a Sub-Program of Classical Liberalism

It was Milton Friedman, one of the founders of the Chicago school as a hub of neoliberalism, who launched monetarism as the anti-Keynesian policy program at a time when the policy sphere was ruled by Keynesianism I. Neoliberalism is essentially a sub-program of classical liberalism that attempts to restore liberalism in the classical sense through modernized policy strategies. It has created a global policy trend towards reducing the role of the government and increasing that of the market, especially as from the 1980s; market-oriented economic reforms, such as deregulation and privatization of public enterprises, have been extensively promoted.[5]

While neoliberalism has made institutional reforms in the microeconomic area of the market a major policy issue, monetarism constructed a similar policy strategy, based on classical liberalism, in the area of macroeconomic policy. In fact, monetarism and neoliberalism share the core tenet of classical liberalism that the role of governments should be as limited as possible. Naturally, Keynesianism clashed with them because it originally emerged as a critique of classical liberalism. This is exactly the reason why monetarism is seen as a "counter-revolution" against the Keynesian Revolution.

The shift from Keynesianism to monetarism in the real world came about in a very dramatic way. The birth of the Thatcher administration in the UK marked the beginning of the global conservatist revolution. Peter Hall, using his own concept of "policy paradigm", described the situation thus:

---

[5]  Authors define the term "neoliberalism" in many ways. Sometimes, it includes not only market liberalism or structural reform but also macroeconomic austerity policy. This study considers the neoliberalism and the macroeconomic austerity of the Austrian school as separate policy programs, although they share the hard core of classical liberalism.

Keynesians tended to regard the private economy as unstable and in need of government intervention; monetarists saw the private economy as basically stable and government intervention as likely to do more harm than good. Keynesians saw unemployment as a problem of insufficient aggregate demand, while monetarists believed that a "natural" rate of unemployment was fixed by structural conditions in the labor market that would be relatively impervious to reflationary policy. Keynesians regarded inflation as a problem arising from excess demand or undue wage pressures that might be addressed by an incomes policy; monetarists argued that inflation was invariably a monetary phenomenon containable only by controlling the money supply.

In the 1970s and 1980s, then, Britain witnessed a shift in the basic policy paradigm guiding economic management. Thatcher's policies were not simply ad hoc adjustments to pieces of policy; they were rooted in a coherent vision associated with monetarist economics. Today mainstream economics has synthesized portions of both the monetarist and Keynesian paradigms. In the 1970s, however, two competing doctrines contended for control over British policy, and the monetarist paradigm emerged victorious. (Hall 1992, 92)

Hall's understanding of Keynesianism and monetarism as two competing policy paradigms can be further clarified by using Imre Lakatos' concept of "hard core" and "protective belt" for the two policy programs.

As Hall says, Keynesians generally consider the market economy to be inherently unstable, while monetarists see it as stable. Thus Keynesians believe that the economy sometimes needs government intervention for stabilization. In contrast, monetarists hold that the economy is destabilized if there is arbitrary intervention by the government. These opposing worldviews of Keynesianism and monetarism are the "unfalsifiable hard core" of each policy program. They basically fall into the category of ideological "isms", which are by no means falsifiable scientific propositions. After all, every economic policy, including Keynesianism and monetarism, has emerged from these unfalsifiable metaphysical worldviews and the value judgments linked to them.

On the other hand, in the protective belt of each policy program, there is a policy strategy with a certain theoretical and empirical basis. As mentioned above, early Keynesians saw fiscal policy as the main policy tool for overcoming recession; to this, the fiscal multiplier theory provided support. Around the end of the 1960s, the focus of macroeconomic policy shifted to overcoming inflation, rather than recession. Some Keynesians then began to propose a wage control policy called "income policy" to contain inflation.

Monetarists criticized this Keynesianism I policy strategy, and proposed an opposing policy strategy, namely rule-based monetary policy. Unlike Keynesians, monetarists do not think that recession is inevitable in the market economy. They believe that the market economy has an inherent ability to stabilize itself. According to them, the occurrence of a serious recession is the result of government intervention distorting the normal operation of the market.

A typical example of such thinking can be seen in *A Monetary History of the United States; 1867-1960* (Friedman and Schwartz 1963). In Chapter 7, titled "The Great Contraction, 1929–33," Friedman and Schwartz argue that the Great Depression was simply caused by a failure of the US Federal Reserve's monetary policy. This contention virtually denies the Keynesian notion that the Great Depression was essentially a manifestation of the intrinsic instability of the market economy.

The macroeconomic policy strategy derived from such thinking was rule-based monetary policy. According to monetarists, the market economy is basically stable, so there is no need to stabilize it by fiscal policy. The problem lies in monetary policy. In the first place, it is impossible for anyone other than the monetary authorities, i.e. the government or the central bank, to supply the market with the money that people need. Monetary authorities can either increase or decrease the money supply. If the authorities increase the money supply, then its value declines and inflation occurs. If they reduce the supply, deflation occurs, which, in turn, leads to a stagnant economy. According to monetarists, the main cause of macroeconomic instability is arbitrary monetary policy management by the monetary authorities. This leads them to their conclusion that it is necessary to impose a monetary policy rule on the monetary authorities instead of granting them discretion.

2. Natural Rate of Unemployment Hypothesis and its Policy Implications

It is Friedman's natural rate of unemployment hypothesis that theoretically supports the monetarist policy strategy. The starting point was a critique of the traditional interpretation of the Phillips curve, proposed in his 1968 presidential address at the American Economic Association (Friedman 1968). Friedman then argued that the trade-off between inflation and unemployment, as described by the Phillips curve, only holds in the short run, when workers do not expect inflation; it does not hold in the long run, when inflation is factored into their expectation.

Friedman assumes that there is a steady unemployment rate that is unrelated to the level of inflation, just as there is a neutral interest rate that does not cause inflation or deflation in the economy. Following Knut Wicksell, a Swedish monetary theorist, who named the neutral interest rate "the natural rate of interest", Friedman called the steady unemployment rate of the economy as the natural rate of unemployment.

Friedman argues that the actual rate of unemployment deviates from the natural rate because workers do not expect actual price changes and, thus, the rate of inflation people expect is different from the actual inflation rate. This is the reason why there is a trade-off between the actual inflation and the unemployment rate in the short run. However, if the actual inflation rate is incorporated into the inflation rate the people expect, the short-run

Phillips curve itself shifts, and the actual rate of unemployment returns to the initial natural rate. Therefore, the long-run Phillips curve becomes a vertical line at the level of the natural rate of unemployment.

The monetarist policy strategy of making the monetary authorities follow a strict money supply rule is based on this natural rate of employment hypothesis. According to this hypothesis, the change in employment, that is, the divergence of the actual unemployment rate from the natural rate, is caused by the expectation error, which is the deviation of the expected inflation rate from the actual inflation rate. Therefore, it is important for employment stability to stabilize inflation at a certain level and to make the inflation rate the people expect match the actual inflation rate. To that end, it is desirable for the monetary authorities to increase the money supply at a fixed rate, such as k %, rather than increase or decrease the money supply arbitrarily. This is the "k % rule," which was once synonymous with the monetarist policy strategy.

3. Adaptation of Monetarism: From the Natural Rate to NAIRU

Monetarism had been making rapid inroads into both academia and the world of policy makers since the late 1960s as inflation and stagflation became a global issue. In the academic world, the ideas of the monetarist counter-revolution were inherited by the rational expectation revolution, which evolved into the new classical macroeconomics. Keynesian economics had been supposed to have been eliminated from the forefront of macroeconomics. It was at that time that Robert Lucas proclaimed "the death of Keynesian economics".

However, monetarism could not ensure its own survival as a policy program. It certainly had a profound effect on macroeconomic policy during the two conservative regimes established around 1980, that is, the Thatcher and Reagan administrations. However, this was not only the first, but also the last time that monetarism affected actual macroeconomic policy. Since then, economists recognizing themselves as monetarists have dwindled in both academia and the world of policy.

In contrast, Keynesianism as a policy framework has made its way through heavy critical fire and once again regained its earlier status as a driver of macroeconomic policy. This is because Keynesianism was able to evolve its policy strategy by heeding certain criticisms by the anti-Keynesian classical economists and incorporating their theories and policy ideas.

A typical example of this Keynesian adaptation in response to critics' views is the non-accelerating inflation rate of unemployment (NAIRU) hypothesis proposed by Franco Modigliani and Lucas Papademos (1975).[6]

---

[6] NAIRU was originally named NIRU in Modigliani and Papademos (1975).

At first glance, NAIRU looks similar to the natural rate of unemployment. However, the logic behind it is very different, as shown by Tobin (1999). In Friedman's view, a trade-off between inflation and unemployment occurs only when there is an expectation error. If it does not exist, a vertical long-run Phillips curve is established. The unemployment rate at that point is the natural rate of unemployment, as defined by Friedman.

The NAIRU hypothesis, on the other hand, presumes that a Phillips curve has a near horizontal area and a near vertical area, depending on the size of the output gap. When the output gap is large, there might be a large amount of idle labor in the labor market. Therefore, the market impact of raising nominal wages and prices would not be so large even if the unemployment rate fell. When the output gap exceeds zero, however, the labor market could become tight owing to scarcity of idle labor. In such a situation, a minimal decline in the unemployment rate would see nominal wages and prices accelerating rapidly. NAIRU is the unemployment rate that is established exactly at this boundary between the horizontal area and the vertical area of the Phillips curve.

The natural rate of unemployment hypothesis and the NAIRU hypothesis also differ in their policy implications. The primary condition for achieving the natural rate of unemployment is to make the expected inflation match the actual inflation, which is why monetarism proposed a rule-based monetary policy.

In contrast, the NAIRU hypothesis has the following two policy implications. First, it suggests that the Keynesian policy for aggregate demand expansion accelerates the inflation rate in the vertical phase of Phillips curve; thus, it has little effect on reducing the unemployment rate. Second, it suggests that the Keynesian policy could bring about substantial improvement in employment and income without substantial inflation in the horizontal phase of the Phillips curve.

The NAIRU hypothesis can be interpreted as a Keynesian adaptation of the natural unemployment rate hypothesis — a reconstruction of it as a theory protecting the hard core of Keynesianism. It partially accepts Friedman's contention that a vertical Phillips curve would hold if the unemployment rate reached its structural level. On the other hand, it reveals that macroeconomic policy could be useful as long as the unemployment rate has not reached that structural level; further, NAIRU is the limit up to which it could be useful. In that sense, the NAIRU hypothesis played a role in protecting the core notion of Keynesianism that has focused on employment and income stabilization with the application of macroeconomic policy.

4. Transformation of Keynesianism Triggered by New Classical Macroeconomics

In the 1960s and 1970s, when Keynesianism directly confronted monetarism, an understanding common to both policy programs was that Keynesianism aimed to realize employment expansion through fiscal policy, while monetarism sought to realize the stabilization of general prices through monetary policy. In the wake of criticism by proponents of monetarism, however, Keynesianism began to reformulate the policy strategy as part of its protective belt. While maintaining the core value judgment that the economy always needs macroeconomic stability, more emphasis was placed on stabilization of the inflation rate. Further, the focus was brought to bear more on monetary policy than on fiscal policy. This adds up to Keynesianism II.

This new direction of Keynesian policy strategy was, in part, a result of what was going on in the academic world. It saw the rise of a new Keynesian economics that was developed after borrowing some notions from the new classical macroeconomics. Thus, Gregory Mankiw and David Romer, who represent the new Keynesians, stated that "much of new Keynesian economics could also be called new monetarist economics" (1991, 3).

Keynesians and monetarists were initially in conflict over the "discretion or rule" needed to conduct macroeconomic policy. Keynesians considered that the government could achieve nothing without using discretion in its policy decisions to stabilize the volatile market. Monetarists thought that such arbitrary policy decisions of the government were the primary reason for market instability. This conflict was resolved after the rational expectations theory, which followed monetarism and clarified the role of "expectations" in macroeconomic policy. However, this was not because Keynesians were convinced by the new classical macroeconomists' doctrine that stabilizing expectations would naturally stabilize the economy. Instead, Keynesians realized that some control of people's expectations is also necessary to make macroeconomic policy effective.

The vital difference between discretion and rule in macroeconomic policy lies in the predictability of the policy. In the case of discretionary policies, people cannot accurately predict them. If policies are set in advance as rules, people can foresee, to some extent, what the government will do. Further, if the government carries out ad hoc policies that are not based on rules, their results will be less predictable, which, in turn, will make them less effective.[7] This means that government policies must not be unpredictable, even if the market economy itself is unforeseeable.

However, Keynesians and the new classical macroeconomists have basically different worldviews on the economy. Thus, even if the two groups superficially appear to be close, there will always remain essential differences between the two policy strategies. For example, there is an apparent difference in the way these two camps grasp policy rules.

First of all, monetary policy rules in Keynesianism cannot be as mechanical as the k% rule in monetarism. In monetarism, a mechanical monetary rule of the kind is both necessary and sufficient for macroeconomic stabilization; this is because the central bank's arbitrary monetary policy is the main source of destabilization in the economy. In contrast, Keynesians hold the view that the market economy has an inherently unstable characteristic

---

[7] These problems are generally referred to as dynamic inconsistency, the notion first presented by Kydland

because the investment demand of enterprises as well as the aggregate demand of the economy as a whole is based on entrepreneurs' capricious investment decisions. Therefore, it is necessary for central banks to embark on active policies that offset fluctuations in private demand, rather than sticking to conservative policy rules. In other words, the central banks have to expand the money supply more vigorously when private demand is shrinking; similarly, they have to reduce it when the economy is overheating owing to the expansion of private demand. Thus, the rule that is needed is not a mechanical one, such as the k% rule, which pays no attention to the prevailing economic conditions, but a state-dependent policy rule that adjusts policies according to them.

It was John Taylor who formulated a Keynesian-type monetary policy rule (Taylor 1993). According to the Taylor rule, as it is called, the central bank decides the policy interest rate on the basis of two variables indicating the macroeconomic situation; these are "the size of the output gap" and "the deviation of the inflation rate from its target level". Output gap and inflation rate are the most basic economic indicators that show whether the economy is stagnating or overheating. Further, the policy rate is the most basic policy instrument that central banks use to ease or tighten money. This means that the Taylor rule is precisely the Keynesian monetary policy rule that central banks adjust monetary policy according to existing macroeconomic conditions.

With the development of new Keynesian economics, Keynesian policy strategy has evolved to a monetary policy based on the framework of inflation targeting. Macroeconomic stability in Keynesianism means achieving the potentially attainable income and employment by bringing the output gap as close to zero as possible; at the same time, price stability is realized by stabilizing the inflation rate. This is equivalent to achieving the lowest unemployment rate that does not accelerate inflation, or in other words NAIRU. Further, it is also equivalent to achieving the lowest inflation rate that does not further expand unemployment.[8] Because of this, the major central banks have usually set their target inflation rate at around 2%.

The basic idea behind inflation targeting is constrained discretion, which is another expression of the state-dependent flexible rule described above. Inflation targeting requires central banks to achieve and maintain the target inflation rate. However, it does not impose mechanical rules on the policy operation. In order to achieve and maintain the target inflation rate, the central bank needs to adjust its policy interest rate according to the output gap and inflation trends, as the Taylor rule suggests. However, the decision is left to the discretion of central banks. This is what is meant by constrained discretion.

Inflation targeting is the new protective belt in Keynesianism II that was triggered by the monetarist counter-revolution and the rational expectations revolution. It was the most and Prescott (1977).

[8] A theoretical foundation for the view that the optimal inflation rate is not 0%, but a positive value around 2% was presented by Akerlof, Dickens, and Perry (1996).

refined form of policy strategy in Keynesianism II at this stage.

## V. The Great Recession and New Policy Strategy of Keynesianism II

1. Changes of Macroeconomic Policy during the Great Recession

Keynesianism II, which follows the traditional monetary policy of adjusting the policy interest rate according to the economic situation, enjoyed its heyday during the so-called Great Moderation. This was the period from the mid-1980s to the 2008 financial crisis; the term became popular after Ben Bernanke, then a governor of the US Federal Reserve, used it in a speech (Bernanke 2004). During that period, economic growth with moderate inflation and relatively small fluctuations in the economy was achieved in many developed economies. As Bernanke stated in the speech, monetary policy was apparently one of the key factors that brought about such moderation in the global economy.

The global financial crisis that began in September 2008 and the subsequent global downturn — called the Great Recession — changed the situation. Since then, the policy interest rate has fallen to the lowest bound in many countries. This means that the central banks lost the room for implementing traditional monetary policy. As a result, some central banks have shifted to non-traditional monetary policies, such as quantitative easing. In the fiscal sphere, many countries initially implemented expansionary fiscal policies, in a throwback to Keynesianism I. However, in the wake of the Greek crisis in the spring of 2010 and the subsequent sovereign debt crisis of the Eurozone countries, these expansionary fiscal policies were abandoned in rapid succession. Instead, fiscal austerity was adopted worldwide.

This austerity policy, as a backlash against Keynesian fiscal expansion, led to serious delays in the recovery of the global economy. The major victims of austerity were the Eurozone debt-crisis countries, such as Greece, Spain, and Portugal, where unemployment increased rapidly, especially among the youth. This economic stagnation caused by austerity policies, and the resulting desolation of the economies, have since gone under the epithet of "austerity fatigue".

The worldwide "anti-austerity" tide arose under such circumstances. The movement is very complicated, transcending political groups on the right and left. However, it has a common economic policy slogan, namely, anti-austerity.

2. From Traditional to Non-Traditional Monetary Policy

Traditional monetary policy and non-traditional monetary policy are distinguished by the fact that central banks do not use the policy interest rate as the operational target of monetary policy in the latter. Traditional monetary policy seeks to stabilize employment, prices and income by manipulating the policy interest rate. In contrast, non-traditional monetary policy uses means other than policy interest rates to achieve the same purpose.

As mentioned above, the financial crisis that started in the fall of 2008 and the subsequent global recession forced the major central banks to embark on massive monetary easing to restore stability in the financial market and prevent a further dip in the economy. As a result, some central banks fell into a "liquidity trap", that is, the lowest bound of the policy interest rate, wherein traditional monetary policy can no longer work. Non-traditional monetary policy was a natural outcome.

It was Ben Bernanke who played the most important role in the development of both the theory and practice of non-traditional monetary policy. Prior to joining the Federal Reserve, Bernanke had published a paper titled "Japanese Monetary Policy: A Case of Self-Induced Paralysis?" (Bernanke 2000). In the late 1990s, the Japanese economy had already faced a liquidity trap, long before the developed economies' experience since 2008, owing to the long-term stagnation accompanied by deflation. However, the Bank of Japan chose policy inaction, arguing that monetary policy could do nothing once it reached the lowest bound of the policy interest rate. Criticizing this view, Bernanke argues, "despite the apparent liquidity trap, monetary policymakers retain the power to increase nominal aggregate demand and the price level" (Bernanke 2000, 158).

In his article, Bernanke presented viable monetary policy options under a liquidity trap. The most important of them were "nonstandard open-market operations" by central banks. It was this idea that led to the practice of large-scale asset purchases (LSAPs), commonly referred to as quantitative easing, about ten years after.

Under the liquidity trap, central banks cannot lower the policy interest rate, no matter how much money they pump into the system. However, if the central bank expanded the base money by purchasing "nonstandard" assets that were not substitutes for money, the asset market would definitely be affected. This mechanism is generally called portfolio rebalancing. Underlying the idea is the general equilibrium analysis of the asset market which was proposed by James Tobin (Tobin 1969). In short, the quantitative easing policy affects employment and prices through the rise in asset prices caused by portfolio rebalancing.

Bernanke launched the LSAPs program during the subsequent global economic crisis. As to the policy, he placed the greatest emphasis on this asset channel that is based on portfolio rebalancing. In a speech given at Jackson Hole in August 2012, Bernanke explained the mechanism of the policy as follows.

> Federal Reserve purchases of mortgage-backed securities (MBS), for example, should raise the prices and lower the yields of those securities; moreover, as investors rebalance their portfolios by replacing the MBS sold to the Federal

Reserve with other assets, the prices of the assets they buy should rise and their yields decline as well. Declining yields and rising asset prices ease overall financial conditions and stimulate economic activity through channels similar to those for conventional monetary policy. (Bernanke 2012, 4)

If the central bank purchases various assets owned by private financial institutions and supplies the base money, the price of risk assets, including stocks, rises and the exchange rate depreciates as a result of portfolio rebalancing. A rise in stock prices means that the expected return on investment outweighs the cost of the investment, leading to investment expansion. If the exchange rate of the home currency falls, the production of the export industry and the industry making import substitutes expands. An increase in land prices means an increase in the value of real assets held by companies and households, which contributes to an increase in their expenditure. Therefore, the quantitative easing policy has the effect of stimulating economic activity.

The expansionary effect of monetary policy through the interest rate channel does not work in the liquidity trap. However, the effects through asset prices, such as that of stocks, exchange rates, and land, remain. This is why Bernanke stated that a non-traditional monetary policy is as effective as a conventional monetary policy.

3. New Role of Monetary Policy in Allowable Fiscal Deficit

Keynesianism generally advocates a policy vision wherein governments actively implement counter-cyclical macroeconomic policies to achieve macroeconomic stability of the economy. An important corollary derived from this vision is the notion of allowable fiscal deficit, which states that the government's fiscal balance should be realized through the course of the entire business cycle and not within each accounting period; thus, a temporary deficit during a recession is allowable. This notion has existed as a policy strategy of Keynesianism since it took shape.[9] However, the Greek crisis and the European sovereign debt crisis that followed cast serious doubts on the validity of this notion, for the crisis seemed to have been caused by unhindered deficit expansion during the recession.

This shows that the notion of allowable fiscal deficit was not well-grounded. From a Keynesian perspective, fiscal deficits during recession should be allowed as much as possible. However, once the deficit actually grows, fiscal concern inevitably arises. Further, there is no denying that what began as concern might eventually turn into a real crisis. This is exactly why many countries made what turned out to be the worst choice — austerity during recession.

---

[9] This idea began with Keynes' criticism of the "Treasury View". Lundberg (1985, 8) noted that the idea of a cyclical budget balance was first proposed by Gunnar Myrdal in his Appendix to the Swedish government's fiscal program of January 1933. The notion was further developed by Abba Lerner, as his theory of functional

An expansionary monetary policy was necessary to avoid making such a choice. The following are three measures that would have helped avoid a debt crisis.

The first is stabilization of the government bond market through the purchase of government bonds by the central bank. Monetary easing by the central bank generally involves purchasing government bonds and increasing supply of the local currency. During a recession, government bond issuance usually increases owing to a decrease in tax revenues and the fiscal spending required for economic stimulus. This can sometimes disturb the bond market. However, if the central bank implements monetary easing through the purchase of government bonds, the disturbances in the government bond market can be suppressed.

The second is reduction of government bond interest payments through expansion of the central bank's bond-holding; if the government issues bonds to cover the budget deficit, it must continue to pay interest to government bond-holders in the private sector. However, if the central bank buys the government bonds, the government can escape its interest payment obligations, for the interest payments on government bonds held by the central bank return to the government's coffers. Therefore, if the central banks expand their holdings of government bonds, the government's interest payments to the private sector as a whole necessarily decrease. Because holding debt entails the necessity to pay interest, the government debt held by the central bank is virtually equivalent to the absence of government debt. This is what is known as seigniorage.

The third is the fiscal improvement through economic recovery induced by monetary easing. Monetary easing generally improves employment, income, and corporate earnings through various channels. Because tax revenue is dependent on household income and corporate earnings, the improvement in income and corporate earnings naturally leads to improvement in tax revenue, which had decreased owing to the recession.

In principle, the autonomy of monetary policy means that the government's fiscal balance can be delayed longer than would be possible without autonomy.[10] The European countries, such as Greece, Spain, Portugal, Ireland and Italy, which faced fiscal crisis have forgone monetary autonomy with unification in the euro. Actually, it was probably no accident that all these debt-crisis countries joined the Eurozone. In this respect, the European sovereign debt crisis was not a problem of fiscal discipline, but a problem of the "euro fetters," that is, the result of countries participating in the monetary unification project being deprived of monetary autonomy.

---

finance (Lerner 1943).

[10]  Modern Monetary Theory (MMT), proposed around the middle of the 1990s, representatively by Warren Mosler and Randall Wray, maintains that there is no government fiscal constraint in an economy with a sovereign currency. See Wray (2015).

## 4. Friedman-Bernanke Helicopter Money

The combination of a policy of monetary easing and deficit fiscal policy is not new in itself. It is a policy that has been discussed as "money finance", or "helicopter money". The novelty is that this integrated strategy is positioned as a realistic policy plan against austerity, and not just as a fable in economics.

Bernanke, the then Fed governor, gave a talk in November 2002 entitled "Deflation: Making Sure 'It' Doesn't Happen Here" (Bernanke 2002). The background to this talk was the US economy in recession at the time owing to the collapse of the IT dotcom bubble, the inflation rate dropping sharply. The then Fed chairman, Alan Greenspan, feared that the US would fall into a deflation similar to that in Japan (Greenspan 2007, 228). Were it to happen in the US, Bernanke proposed the following policy option:

> In practice, the effectiveness of anti-deflation policy could be significantly enhanced by cooperation between the monetary and fiscal authorities. A broad-based tax cut, for example, accommodated by a program of open-market purchases to alleviate any tendency for interest rates to increase, would almost certainly be an effective stimulant to consumption and hence to prices. Even if households decided not to increase consumption but instead re-balanced their portfolios by using their extra cash to acquire real and financial assets, the resulting increase in asset values would lower the cost of capital and improve the balance sheet positions of potential borrowers. A money-financed tax cut is essentially equivalent to Milton Friedman's famous "helicopter drop" of money. (Bernanke 2002, 3-4)

This observation earned Bernanke the nickname "Helicopter Ben".

Friedman's helicopter money argument appeared in his article titled "The Optimum Quantity of Money" (Friedman 1969). In order to show what happens when the quantity of money is expanded, Friedman proposed a thought experiment involving a helicopter that dropped banknotes over people's heads. The result is, of course, a rise in prices. Bernanke simply adapted this idea to propose a policy that prevents deflation.

Helicopter money is, in short, the policy to cover government fiscal spending not with government bonds, but with the central bank's money. In order to achieve this, the central bank has to buy the deficit bonds issued by the government and supply the base money to the market. Although the underwriting of government bonds by the central bank is legally prohibited in many countries, the central bank can purchase the same amount of government bonds from private financial institutions. As a result, the money held by private enterprises and households increases by the amount of the base money supplied by the central bank, without decreasing other assets held by the private sector. The result is exactly the same as that achieved by "dropping money from a helicopter".

The helicopter money policy, or money finance, was counted among the most unsound economic policies that would necessarily lead to uncontrollable inflation. It is now seen as a serious policy option for coping with the secular stagnation that has emerged since the Great Recession.[11]

## VI. The Hard Core and Protective Belt of Keynesian Policy Program

Keynesianism is a policy ideology stating that the market economy needs positive counter-cyclical macroeconomic policy to stabilize the inherently unstable market economy. The core of Keynesianism consists in this vision and the related value judgment that macroeconomic stability, especially the stability of income and employment, is important. This core does not change as long as Keynesianism endures.

On the other hand, the protective belt of Keynesianism consists in its policy strategy: specific policy targets and instruments for securing macroeconomic stability. This policy strategy changes with changing realities and the acquisition of newer knowledge. To correctly understand the causal relationship between policy target and its instruments, it is necessary to infer it on the basis of a revised and more reliable scientific theory. However, just as every scientific theory inevitably changes with the progress of theoretical and empirical research, so must policy strategy.

The initial Keynesian policy strategy that relied mostly on fiscal policy lost potency precisely for this reason. It reflected the changes that occurred in the intellectual domain of economics. Nevertheless, Keynesianism as policy program survived. Robert Lucas' prophecy that Keynesianism would die in the policy world, as well as academia, was not fulfilled. Rather, it was monetarism that died out. This is because Keynesianism was able to evolve its policy strategy into something more realistic by incorporating the theoretical tools proposed by its critics.

The vitality present in Keynesianism was never to be found in anti-Keynesian policy programs, such as monetarism and the Austrian school. The differences probably come from the degree of "realism" that these programs incorporate.

At the core of anti-Keynesian programs, such as monetarism and the Austrian School, is the idea inherited from classical liberalism that the market economy is inherently stable, and that the government always tends to destabilize it. Such a view of the world inevitably leads to a policy strategy that tends to narrow the scope of the government's policy as much as possible. Some examples are the k% rule for monetarism and the restoration of the gold standard favored by the Austrian school. However, it became apparent after all that these policies could not achieve the results to which our society aspires.

Keynesianism started with criticism of this worldview of classical liberalism. It sees the market economy as a highly volatile economic system; it is marked by unavoidable uncertainty, as well as income and employment fluctuations, caused by fluctuations in demand. Keynesianism supposes that depression is a manifestation of this essential attribute of the market economy. When such economic turmoil occurs, therefore, Keynesianism supposes that the government as an agent entrusted with the public's interest should definitely intervene in the economy to improve the situation, rather than sit and wait for disaster.

---

[11]  A leading advocate of modern helicopter money policy is Adair Turner. See Turner (2016).

Unfortunately for our society, this vision at the core of Keynesianism has proved to be all too realistic. Depression has always arrived when society forgot about it. And the aspiration to overcome it has remained. On such occasions, our society has had to find some way of coping with the situation. To this end, Keynesianism has proved the only valid approach. And this, indeed, is the most important reason why Keynesianism eventually survived.